# EXPLOITING METAMATERIALS, PLASMONICS AND NANOANTENNAS CONCEPTS IN SILICON PHOTONICS

Francisco J. Rodríguez-Fortuño[1], Alba Espinosa-Soria[2], and Alejandro Martínez[2,*]
1 Department of Physics, King's College London, Strand, London WC2R 2LS, United Kingdom
2 Nanophotonics Technology Center, Universitat Politècnica de València, Valencia 46022, Spain
* Email: amartinez@ntc.upv.es

## Abstract

The interaction of light with subwavelength metallic nano-structures is at the heart of different current scientific hot topics, namely plasmonics, metamaterials and nanoantennas. Research in these disciplines during the last decade has given rise to new, powerful concepts providing an unprecedented degree of control over light manipulation at the nanoscale. However, only recently have these concepts been used to increase the capabilities of light processing in current photonic integrated circuits (PICs), which traditionally rely only on dielectric materials with element sizes larger than the light wavelength. Amongst the different PIC platforms, silicon photonics is expected to become mainstream, since manufacturing using well-established CMOS processes enables the mass production of low-cost PICs. In this review we discuss the benefits of introducing recent concepts arisen from the fields of metamaterials, plasmonics and nanoantennas into a silicon photonics integrated platform. We review existing works in this direction and discuss how this hybrid approach can lead to the improvement of current PICs enabling novel and disruptive applications in photonics.

## 1. Introduction

Plasmonics is the science that studies surface plasmon (SP) propagation along metal surfaces at visible and near-infrared (NIR) wavelengths [1]. A key property enabled by SPs is the strong light confinement well beyond the diffraction limit in metallic nanostructures via excitation of localized surface plasmon resonances (LSPRs) [2]. This property can be used to build subwavelength-sized plasmonic nanoresonators extremely sensitive to tiny variations in their surroundings (useful in sensing) [3] or plasmonic nanoantennas that efficiently convert confined into radiative fields and vice versa [4]. Isolated metallic nanostructures can also be arranged into 2D or 3D artificial structures named metasurfaces and metamaterials, respectively, which can provide a number of intriguing phenomena such as negative refraction [5], cloaking [6], optical magnetism [7] or phase manipulation [8–10]. Such amazing abilities for manipulating light at the nanoscale have pushed plasmonics, nanoantennas and metamaterials to become research hot-topics nowadays, being strongly related by the fact that their properties ultimately rely on our ability to tailor the optical response (scattering, absorption, polarization, etc.) of subwavelength metallic nanostructures - either isolated [3,7,11–14] or forming compact arrays [15,16] - at will.

The use of metals at optical frequencies comes at a price: high absorption losses prevent efficient waveguiding at long distances (cm-scale). Another important drawback resulting from the subwavelength size of metallic nanostructures is that characterizing them when isolated becomes challenging. Indeed, simultaneous but independent excitation and measurement of multiple nanostructures becomes unfeasible, which is a huge limitation in applications requiring multiplexing such as, for instance, biosensing. Finally, typical metals - mainly gold (Au) and silver

(Ag) - and some of the fabrication processes (such as focused ion beam) are not compatible with standard semiconductor fabrication techniques, mainly CMOS, which prevents bringing plasmonics out of the lab [17].

In parallel, recent years have witnessed the success of silicon photonics technology, from the first experimental demonstrations in the 90s to becoming the mainstream technology for photonic integrated circuits (PICs). The key advantage of silicon photonics over competing approaches (such as III-V semiconductor heterostructures) is the feasibility for mass-scale production of low-cost chips integrating photonic and electronic components made by existing semiconductor fabrication techniques (such as CMOS) [18–21]. Silicon photonics is now a mature technology and silicon PICs fabricated on silicon-on-insulator (SOI) wafers are currently being exploited for label-free biosensors [22–24] and high-speed data communications [25,26] amongst other applications, even with commercial products in the market [27,28].

Some of the properties inherent to silicon PICs would enable overcoming the existing drawbacks related to plasmonics, nanoantennas and metamaterials introduced above. In silicon PICs, light propagates along low-loss silicon waveguides at near-infrared (NIR) wavelengths or along silicon nitride ($Si_3N_4$) waveguides at visible wavelengths. The transverse size of these waveguides is diffraction limited to about $(\lambda/2n)^2$, being $n$ the refractive index of the waveguide core. This means that the confinement is not deeply subwavelength, as plasmonic waveguides may allow, but in contrast they exhibit propagation losses below 1 dB/cm [29], which enables interconnection of processing elements well-spaced (cm-scale or beyond) on a PIC. In this sense, we could think about a hybrid architecture in which dielectric waveguides are used for guiding signals while plasmonic, nanoantenna or metamaterial elements, with their novel features, realize several functionalities (see Fig. 1). This hybrid approach also permits the parallel excitation and measurement of multiple nanostructures by coupling them to several silicon waveguides, which can all be fed simultaneously but detected independently. Finally, realization of plasmonic, nanoantennas or metamaterial-based devices on silicon technology could pave the way towards the industrialization of such technologies. To this end, the use of silicon-compatible materials and processes becomes mandatory.

Silicon PICs could also benefit from the introduction of metamaterials, nanoantennas and plasmonic elements. For instance, the diffraction limit prevents further miniaturization of dielectric circuits below the light wavelength [30], which could be highly alleviated by employing deep-subwavelength plasmonic-based building blocks [2,31]. The relatively weak and slow nonlinear effects in silicon result in large structures operated at high energies for all-optical processing, with speeds limited to some tens of GHz [32]. Giant and ultrafast nonlinearities exhibited by plasmonic nanostructures [33,34] and metamaterials [35] could provide a way towards realizing THz-speed all-optical switching on silicon PICs. Sensitivity to the environment provided by LSPRs in metals surpass that exhibited by dielectric structures, so the performance of silicon photonic biosensors could also be improved by using plasmonics. Finally, nanoantennas could be employed as ultra-compact elements for on-chip and inter-chip wireless interconnects [36] as well as for imaging purposes. Hence, it is clear that enormous benefits may arise from the use of plasmonics, nanoantennas and metamaterials on silicon PICs.

In this paper, we review recent advances showing how new concepts arising from plasmonics, metamaterials and nanoantennas research can be introduced on a silicon PIC. In Section 2, we discuss how single plasmonic nanoresonators can be efficiently coupled to silicon waveguides. This approach can bring huge advantages: for silicon photonics, since it would enable subwavelength components to be used in sensing or switching, therefore enabling denser integration; for plasmonics, since this approach would enable scanning multiple isolated plasmonic nanostructures in parallel and in real-time. In section 3, we discuss recent implementations of metallic as well as dielectric nanoantennas on silicon photonics. Remarkably, we show that, by means of spin-orbit interaction, such nanoantennas enable polarization manipulation at the nanoscale in both transmission and reception schemes, which could be used for building polarization-selective elements on silicon PICs. In Section 4, we discuss the advantages of realizing metamaterials in an on-chip configuration, reviewing recent findings related to the realization of carpet cloaks, hyperbolic metamaterials and epsilon-near-zero metamaterials on a silicon PIC. Main issues related to the technological requirements for exploiting such concepts in a silicon-compatible platform are briefly discussed in section 5.

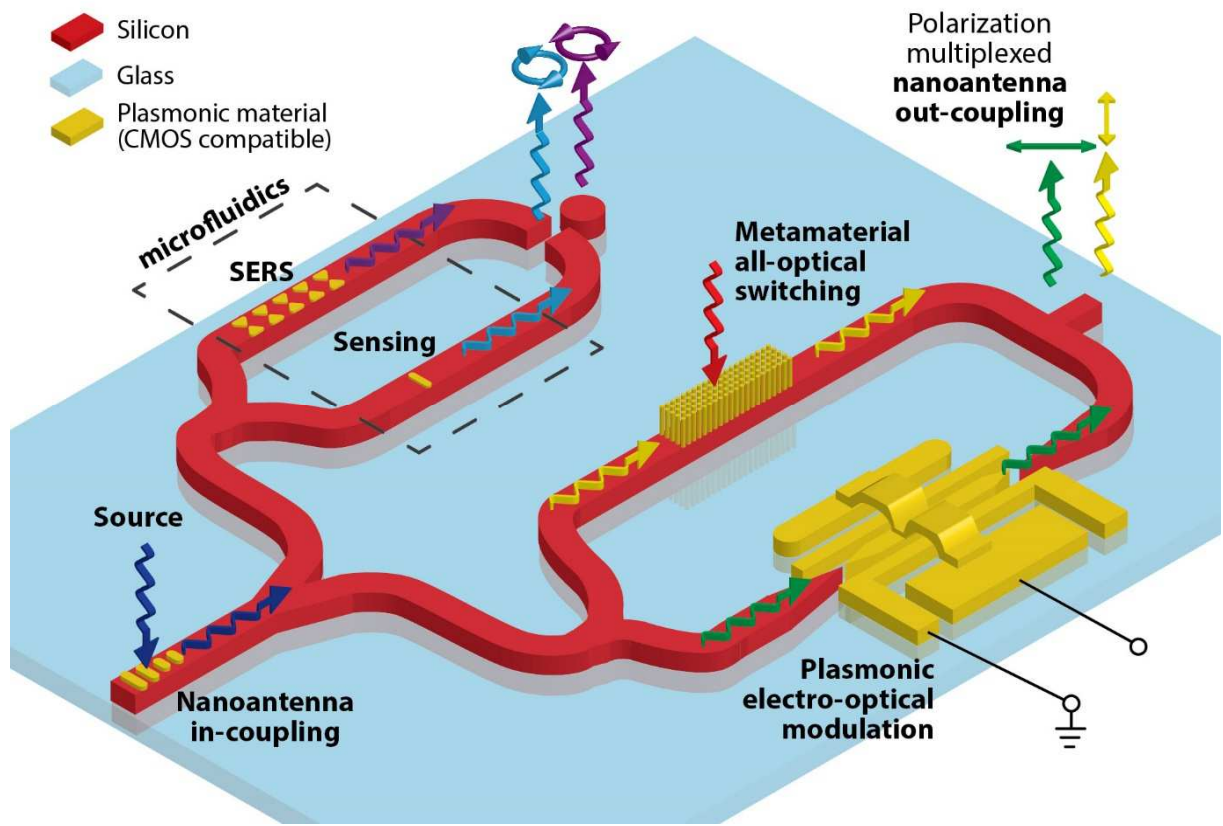

**Figure 1. Scheme of the hybrid architecture for silicon PICs.** Silicon waveguides with negligible losses interconnect compact and efficient plasmonic and metamaterial elements. The left-part of the silicon PICs includes plasmonic elements for enhanced Raman [37] and refractomeric biosensing [38] of a certain fluid. The right-part of the PIC includes electro-optical (with a plasmonic slot Mach-Zehnder interferometer [39]) and all-optical (with a rod-based hyperbolic metamaterial [40]) modulation/switching. Nanoantennas are used as transducer between the outer world and the waveguides: the input port includes a highly directive Yagi-Uda for highly-directional coupling into the PIC and polarization-dependent output nanoantennas for polarization multiplexing of the output radiated light.

## 2. Coupling plasmonic nanoresonators and waveguides to silicon waveguides

In the hybrid architecture previously described, we have a silicon PIC with low-loss silicon (or silicon nitride) waveguides to transport light between different regions of the chip and plasmonic, nanoantennas or metamaterial elements in which the light is processed (for example, biosensing or switching). Such elements should be placed in close proximity to the waveguides to ensure efficient interaction between them and the guided light. In non-guided configurations in which light coming from free-space impinges out-of-plane on a plasmonic element (where out-of-plane refers to incident light from outside the chip) we can assume that the excitation is a plane wave with an electric field that defines the polarization. Indeed, we can adequately choose this electric field as to excite or not a certain LSPR. However, the situation becomes very different when the excitation (and also the collection of the light scattered by the plasmonic element) is realized via high-index waveguides. In this case, the guided light can no longer be approximated by a plane-wave because of the strong confinement by the large index contrast. Indeed, the guided modes show a large longitudinal component of the longitudinal electric field accompanied by a transverse spin [41], which does not exist in plane-wave-like excitation. Figure 2 shows the main transverse ($E_x$ for TE-like and $E_y$ for TM-like) and longitudinal ($jE_z$) electric field components as well as the local helicity of the first two modes of a strip silicon waveguide at $\lambda=1,550$ nm [42]. If we consider the TE-like mode for excitation of a plasmonic nanostructure, only if it is very small (100 nm or less) and symmetrically placed with respect to the $yz$-plane, the longitudinal component vanishes and the excitation can be considered similar to the case of an incident plane wave. However, as long as the nanostructure gets larger or placed in a region with a non-negligible longitudinal component, the effect of this field component has to be taken into account when considering its excitation from the guided field. Indeed, since $E_x$ and $E_z$ exhibit a $\pi/2$ phase shift between them, such regions display local elliptical polarization (see right column in Fig. 2), which will result in spin-orbit interaction effects [42–44]. This interaction, as well as the resulting interesting possibilities of spin-controlled directional guiding phenomenon, are considered in depth in section 3.

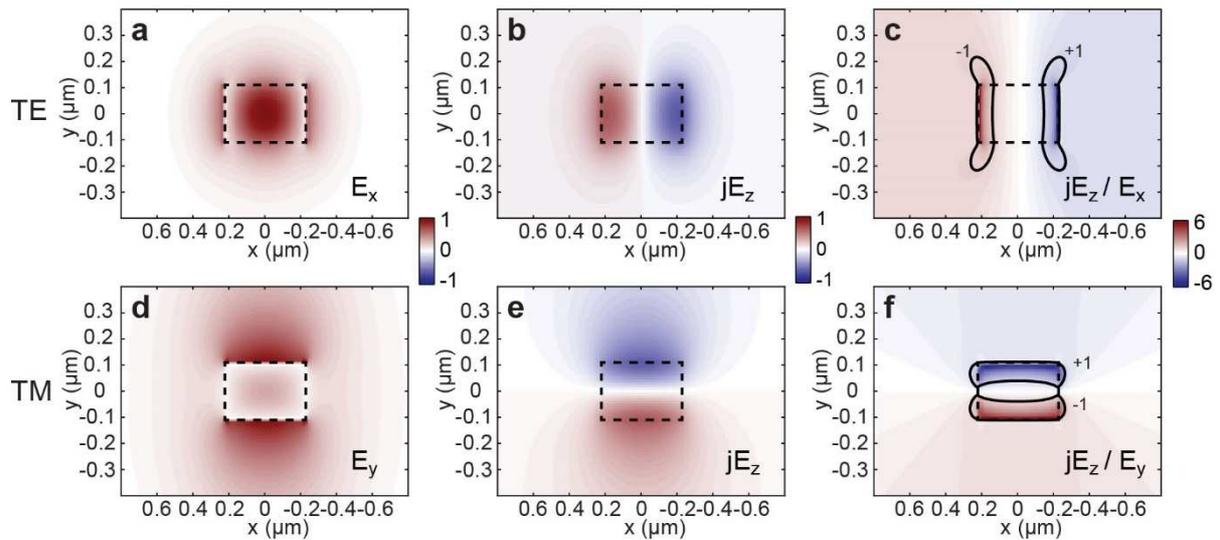

*Figure 2. Electric field components of the two fundamental guided modes (top: TE-like or even mode; bottom: TM-like or odd mode) of a silicon strip waveguide with 400x220nm² rectangular cross-section at λ=1,550 nm. (a,d) Fundamental transverse electric field component responsible for coupling and excitation from external optical fibers; (b,e) Longitudinal electric field component (jEz). (c,f) Local helicity obtained as the ratio between the longitudinal and transverse component; local polarization turns out to be elliptical except for the highlighted contours where it is purely circular. Calculations performed using CST Microwave Studio.*

In this section, we consider small plasmonic nanostructures being symmetric with respect to the $yz$-plane so that the effects related to the longitudinal field component can be disregarded when exciting the LSPR. In particular, we focus our attention on the coupling between an isolated subwavelength metallic nanostructure and a dielectric strip waveguide. This on-chip configuration may be especially relevant for on-chip biosensing, with the important advantage of being able to drive and measure multiple nanostructures coupled to different waveguides in parallel and in real time, even using different wavelengths for each one, which is out of reach in out-of-plane architectures. The simplest way to do so is merely placing the plasmonic element on top of a silicon waveguide. This is precisely what two of the authors did in Ref. [45], where the excitation of LSPRs of coupled aluminium disks on a silicon waveguide at telecom wavelengths was reported (see Fig. 3(a)). Other recent experiments have reported similar results employing different plasmonic nanostructures showing LSPRs in the NIR or visible regimes (in this last case, $Si_3N_4$ waveguides are used), which are characterized by dips in the transmission spectrum [37,46–48,38,49–51]. Since the nanostructure is placed in a region of evanescent field of the guided mode, the interaction between the nanostructure LSPR and the guided field turns out to be weak. Even by delocalizing the guided field by, for example, reducing the waveguide thickness, maximum values of the coupling efficiency (calculated from the extinction ratio, which is obtained as the ratio between the scattered plus the absorbed power to the incident guided field) are relatively small: ~10% in [38], see Figs. 3(c) and 3(d), and ~20% in [50], which means that most of the guided field does not "see" or interact with the plasmonic nanostructure and is therefore not "processed". Such weak interaction can be enhanced by placing several nanostructures on the waveguide [48] [45] (see Fig. 3(a,b)) so that their total cumulative interaction results on observable effects at the output (this is, a sufficiently large extinction ratio). However, by doing so the total length is increased so we miss some of the main features of such nanostructures: the extreme miniaturization to levels beyond the diffraction limit of isolated nanostructures and the possibility to measure their individual response (useful for example in single molecule detection [3]). Therefore, it would be highly desirable to find other ways to place a single nanostructure in order to increase the interaction between the nanostructure resonance and the guided field.

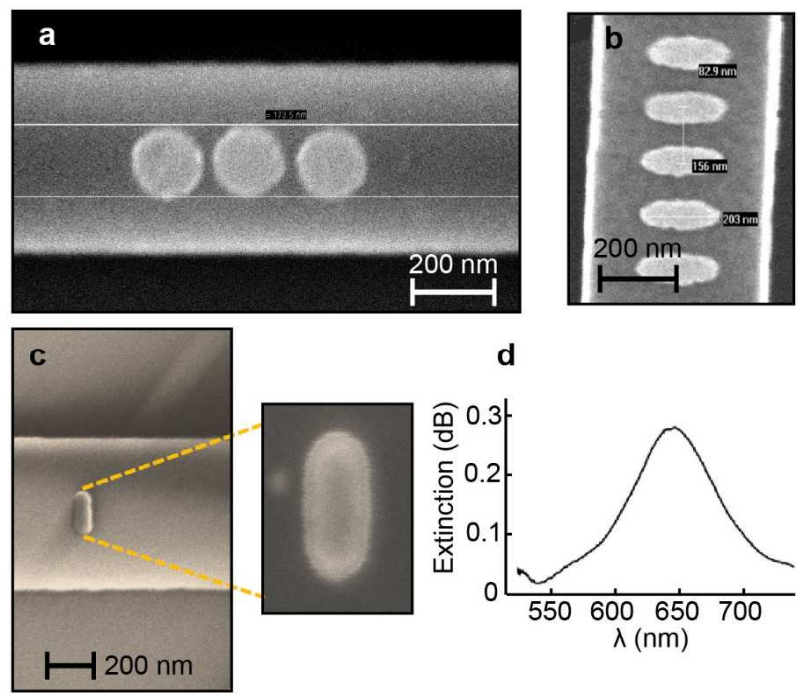

*Figure 3. Plasmonic nanostructures on top of high-index dielectric waveguides:* SEM images of (a) Al nanodisks [45] and (b) Au nanoellipses [48] on a silicon waveguide; (c) A single plasmonic nanoresonator on a $Si_3N_4$ waveguide provides ~10% coupling as measured by the extinction efficiency (scattering + absorption), as shown in the spectrum depicted in (d) [38].

In high-index waveguides, the maximum field intensity of the guided wave is located inside the waveguide core. Therefore, embedding the plasmonic nanostructure inside the waveguide core at the right height would ensure maximum interaction, as recently proposed and numerically analysed

[52]. This approach achieves a strong interaction, but requires digging a relatively deep hole with identical shape and size as that of the plasmonic nanostructure, which may become challenging mainly for nanostructures with complex shapes. An alternative choice that some of the authors proposed and demonstrated experimentally in [53] is to create a subwavelength gap in the waveguide core and place the plasmonic nanostructure inside it, at a height ensuring maximum interaction (see Fig. 4(a)). The field components in the gap region are quite similar to those in a perfect waveguide [53]. This means that if the nanostructure is sufficiently wide, interaction with the longitudinal components of the guided field must be taken into account. However, for sufficiently narrow nanostructures, we can assume that the interaction is governed by the main transverse components of the guided field, so the excitation would become similar to the case of free-space light incidence.

For the simplest case of a gold nanostrip exhibiting its fundamental electric dipolar LSPR at telecom wavelengths, it was observed that the response of the system is mainly characterized by a crossing of the transmission (growing with increasing wavelength) and the reflection (decreasing with increasing wavelength) spectra just around the LSPR wavelength (see Fig. 4(b)). This result can be surprising, since one may expect a dip in both transmission and reflection in such region as a result of the large scattering and absorption of the nanostructure. In contrast, it is the interference between the transmitted and reflected guided fields and the forward and backward scattering produced by the nanostructure what seems to dominate the total optical response. Considering the transmission response, we see two main features. First, at wavelengths below resonance, interference is destructive resulting in low transmission. Indeed, simulation results show that transmission can be reduced by 4-5 orders of magnitude [53], far beyond what can be achieved using isolated nanostructures on top of the waveguides, thus providing a very large extinction ratio at the output, which is key for many applications (sensing, switching). Second, at wavelengths over the resonance, interference is constructive, resulting in a total transmission that becomes even larger than for the case of the gapped waveguide without resonant nanostructure. This can be explained by considering that the nanostructure redirects part of the field scattered by the gap towards the output waveguide. Figures 3(c) and 3(d) show a scanning electron image of a fabricated sample and the measured transmission spectra for both TE-like and TM-like input modes normalized with respect to that of a gapped waveguide without nanostructures. Main features predicted by simulations such as the growing transmission for increasing wavelengths or the normalized transmission over 1 above the resonance are observed here. This configuration is particularly interesting because metallic nanostructures with arbitrarily complex shapes (for instance, those providing magnetic or Fano resonances or even electromagnetically-induced transparency) could be easily inserted in the gap. In addition, the existence of longitudinal components of the field in the gap region give us more possibilities to tailor the nanostructure response. Still, some issues need to be solved. For instance, a perfect alignment between the nanostructure and the waveguide axis is difficult because they are fabricated in different lithography steps, which for some configurations may result in a degraded performance.

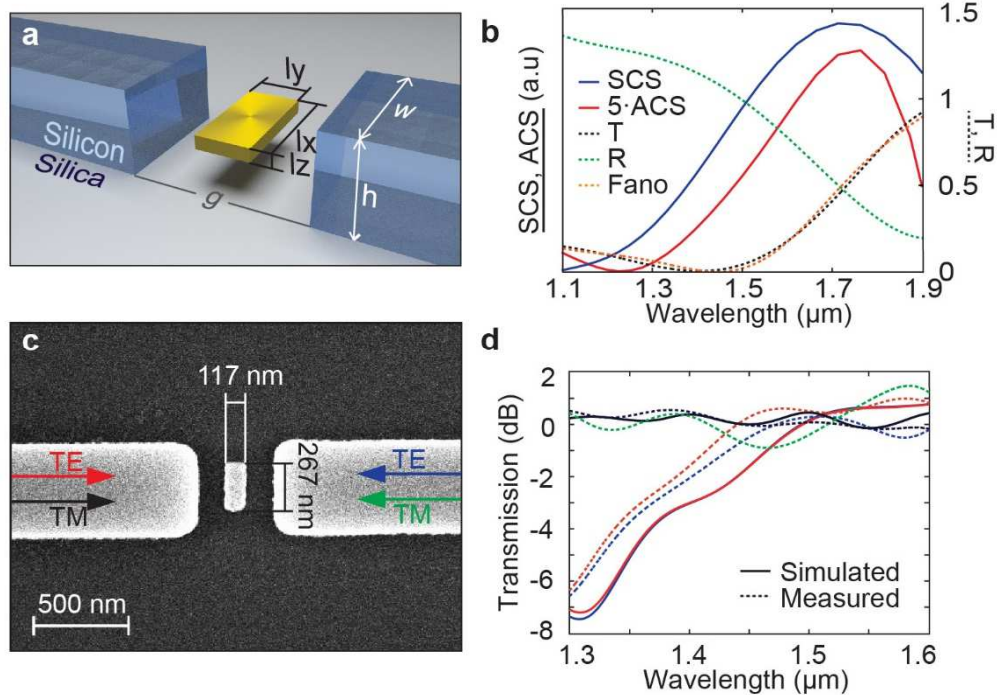

**Figure 4. Plasmonic nanostructure embedded in a Si waveguide gap to maximize interaction** [53]. *(a) Scheme of the proposed structure with a metallic nanostrip placed in the middle of the waveguide gap; (b) Simulation results showing the scattering cross section (SCS) and absorption cross section (ACS) of the considered nanostructure (an Au nanostrip with 320x155x40 nm³) embedded in a silica matrix as well as the transmission and reflection spectra when the nanostructure is embedded in a 400 nm gap created in a silicon strip waveguide. The transmission spectrum can be well fitted using a Fano model, also shown. (c) SEM image and (d) normalized experimental transmission response for TE-like and TM-like input modes of a fabricated sample (the color of the arrows in (c) corresponds to the color of the lines in (d)). A comparison with simulation results is also included. Simulations performed with CST Microwave Studio.*

Besides submicron metallic nanostructures, whether isolated or forming periodic chains, plasmonic waveguides can also be introduced in silicon PICs for compact and efficient photonic processing. In this case, the plasmonic waveguide can be several μm long, so efficient coupling from a silicon waveguide is more straightforward than in the case of isolated nanostructures. For instance, total transfer of optical power from a silicon waveguide to a plasmonic slot waveguide placed on top of it via evanescent coupling has been demonstrated [54]. The plasmonic slot waveguide has received wide attention because of the strong field confinement it can provide in the slot region [55]. Indeed, despite of the larger losses, metallic slot waveguides have shown better performance than their dielectric counterparts [56] in the efficiency of second and third order nonlinear processes, leading to optoelectronic devices with record footprints [57]. Recently, an electro-optical modulator using a plasmonic slot Mach-Zehnder interferometer filled with a nonlinear polymer on a silicon PIC has been demonstrated [39]. In this case the plasmonic waveguides are excited from silicon waveguides using a direct in-line configuration with reduced insertion losses. Therefore, such components – a review on on-chip plasmonic switches can be found in [58] - are highly promising to become building blocks for nonlinear optical processing on a silicon chip. Remarkably, they implement the hybrid architecture described above: silicon waveguides for carrying information at different parts of the chip and nonlinear plasmonic elements with reduced footprint and power consumption for signal processing. Still, some issues mainly concerning with CMOS compatibility need to be addressed, as discussed below in Section 5.

## 3. Dielectric and metallic nanoantennas on silicon photonics

It is well known that, besides providing strong field confinements or large absorption at LSPR frequencies as discussed in Section 2, plasmonic nanostructures can also be used as efficient transducers between nanoscale sources and free-space radiation. In other words, they can bridge near and far fields [4,59,60]. This transducing behaviour resembles that offered by traditional antennas in the radiofrequency (RF) domain –where transmission lines and waveguides can exhibit an extremely subwavelength cross-section– but at scales that are several orders of magnitude smaller [59]. As a result, such plasmonic nanostructures have been termed optical antennas [60] or nanoantennas [61].

So far, many suggested applications of nanoantennas rely on the high intensity of their near fields, known as hot-spots (Fig. 5(b)), achieved under far-field excitation, enabling many different enhanced phenomena such as photodetection [62], gas sensing [63], harmonic generation [11], or vibrational spectrometry [64]. However, here we focus on a more traditional definition of an antenna as that of a device which converts radiated waves into *guided* waves, and vice versa. It is important to notice that this definition, commonly found in electrical engineering textbooks [65,66], does not necessarily involve a subwavelength size in the guided domain (Fig. 5(a)). This way, antennas are used to interconnect wired and wireless networks in an efficient way thus bridging the best of two worlds: signals propagating through free space can simultaneously reach multiple receivers, even if they are at very long distances, without the need for cable deployment; whilst signals in the guided domain can be reliably processed at very high speeds using well-matured technologies.

This traditional concept of antenna as a guided-to-radiated wave transducer can also be applied in the optical domain, and the vast knowledge from RF antenna engineering can be exploited. In the radiated domain, light is used in displays, imaging, communication, and optical manipulation, among many other well-known applications. In the guided domain – for example, in a silicon PIC, as considered in this review -, low-loss dielectric waveguides can be used to carry optical information at different parts of a chip and serve as the basis for information processing at speeds ultimately reaching the THz scale (notice that processes such as demultiplexing [67], switching [68] or amplification [69] can be attained in silicon PICs). The applications enabled by bridging both worlds of radiated and guided light are many. For instance, the possibility of phase tuning the feeding signals of nanoantennas at THz speed would allow exciting applications such as ultra-fast beam-forming and polarization control. Proof of principle demonstrations include arrays of hundreds of dielectric optical nanoantennas used for beam-forming [70], and synthesis of light polarization through the amplitude and phase of the feeding signals on antennas with multiple inputs [71]. Both approaches employ a silicon photonics platform. However, the diffraction limit establishes that dielectric antennas and their feeding waveguides will always have minimum dimensions comparable to the light wavelength. In this sense, dielectric optical antennas are not truly "nano" in scale (Fig. 4c). The relatively large size of dielectric optical antennas results in a significant drawback: when forming arrays, the many-wavelengths period results in unavoidable high-order beams [70]. Alternatively, designing a small dielectric optical antenna comes at the cost of poor transducing efficiency (low effective area) [71]. The diffraction limit in dielectrics therefore imposes a severe trade-off limitation.

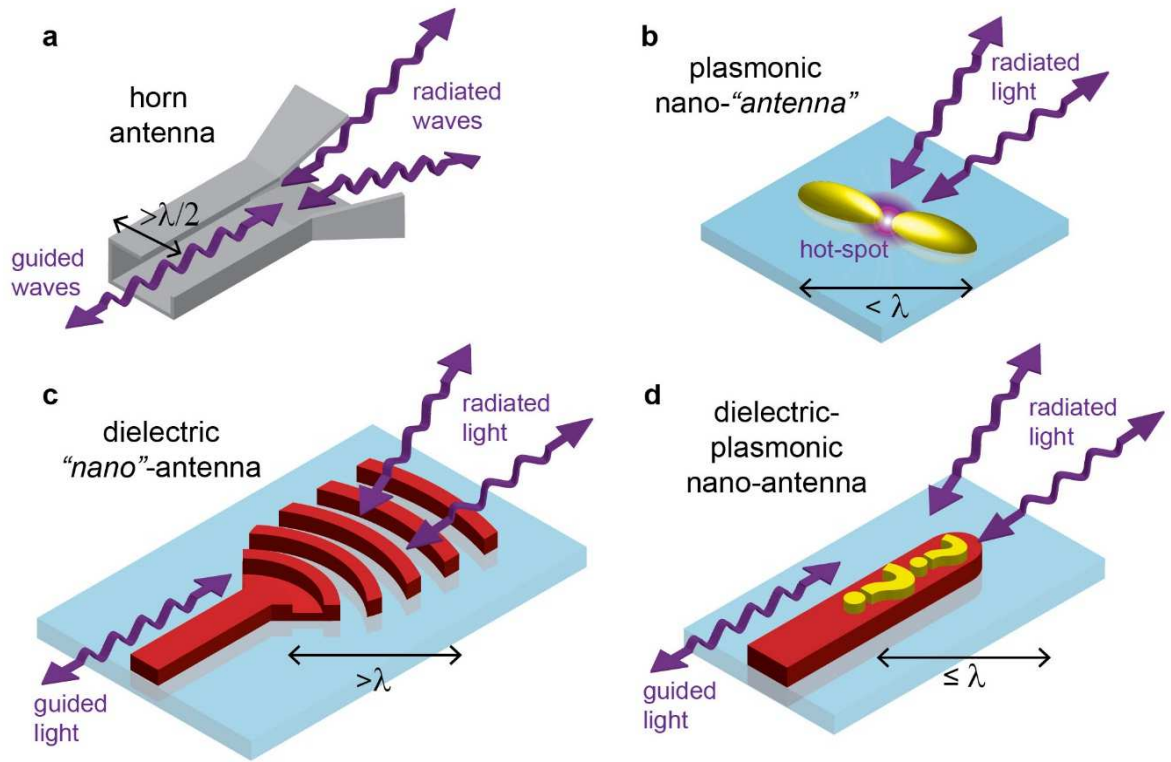

*Figure 5.* **Nanoantennas as a bridge between wired and wireless domains.** (a) A RF horn antenna converts from guided into radiated fields and vice versa in an efficient way. Notice that the waveguide dimensions are comparable to the wavelength; (b) A plasmonic nanoantenna produces a strong subwavelength field spot when illuminated from free-space, and acts as a transducer between a localized point source or absorber and radiated light; (c) A dielectric optical antenna fed by a dielectric waveguide (Ref. [70]) behaves as a classic RF antenna but its size is of the order of -or larger than- the wavelength; (d) Hybrid approach: a subwavelength-sized plasmonic nanoantenna is fed by a dielectric waveguide in order to inherit the best properties of the two worlds. The nanoantenna shape and size should be designed to ensure impedance matching to the waveguide and directional emission in the desired direction.

In order to overcome the diffraction limit and achieve a subwavelength optical nanoantenna without compromising the transducing efficiency, we can turn our attention to plasmonic resonators. This approach, which closely mimics the use of metallic antennas in RF, is similar to that discussed in section 2, but in this case we are interested in the scattering properties of the plasmonic nanostructures. As mentioned above, most plasmonic nanoantennas in the literature are designed to achieve near-field hotspots, with very interesting applications (Fig 4b), but without considering the guided domain. Here we focus instead on those works on plasmonic nanoantennas that follow the traditional definition of antenna and thus include the guided domain –including the use of RF concepts such as input impedances [72,73]. This approach enables exciting plasmonic applications such as wireless optical communications [74] and tunable light beam-forming [75], closely matching those of RF antennas. Beam-forming arrays show no side-lobes thanks to the subwavelength period of the plasmonic nanoantenna elements [75]. Unfortunately, plasmonic waveguides display high propagation losses and, therefore, are currently not generally suitable as long range information highways.

An appropriate solution would be the hybrid architecture as discussed above: using plasmonics resonators as nanoantennas while keeping dielectric waveguides as the feeding elements (or

"wired" network). This brings with it all the advantages of dielectric waveguides named above regarding low-loss and processing abilities, thus gaining the best of each world (Fig. 4d): the subwavelength dimensions and strong field localization of plasmonic nanoantennas with the lossless guiding and fast processing properties of wavelength-size dielectric feeding waveguides. Some preliminary works have shown advances in this direction. For instance, on Ref. [46] a directional plasmonic Yagi-Uda nanoantenna is placed on a dielectric waveguide (Fig. 5a), allowing directional coupling between guided and radiated modes, and on Ref. [76] a plasmonic nanoantenna array on a dielectric waveguide is used for electrically-controlled steering of radiated light (Fig. 5b). Although in the latter case the antenna array does not have a subwavelength period, it has only a few side lobes thanks to the smaller size of the plasmonics elements compared to dielectric nanoantennas.

Yet there is still a lot of work to be performed. For instance, an appropriate matching between dielectric waveguides and plasmonic antennas is mandatory in order to deliver a maximum amount of power to the radiating element whilst minimizing its absorption. The dielectric waveguide could also be modified in such a way that the guided field excites different multipoles in the plasmonic nanoantenna resulting in a highly directional beam, a key property for many applications [13,77,78]. Finally, following the miniaturization trend in photonics, we would like plasmonic nanoantennas to be smaller than the diffraction limit but with an acceptable loss in efficiency, like in the miniaturization of conventional RF antennas. Applying the traditional definition of antennas to the nanoscale, the broad knowledge of antenna theory could lead to unforeseen applications in optics. For instance, we can easily envision a 2D array of subwavelength plasmonic nanoantennas fed by silicon strip waveguides, allowing the efficient conversion of high-bandwidth information-carrying light from a PIC into a free space light beam with high speed tuneable direction, beam shape, and/or polarization.

Another interesting case in which concepts initially developed in plasmonics can be exploited in silicon photonic nanoantennas is the novel idea of evanescent wave spin-controlled directional guiding [41,43,79–82]. This is based on the existence of a large longitudinal component of the electric field of the guided mode, which is antisymmetric with respect to one of the planes bisecting the waveguide (see Fig. 2). As already mentioned, this in general results in an elliptical polarization of the local electric field, whose spinning direction is uniquely locked to the propagation direction of the mode (Fig. 2(c),(f)). This enables spin-orbit interactions when a scatterer is asymmetrically placed in the waveguide proximity, allowing the use of circular polarizations of out-of-plane illuminating light (carrying spin angular momentum) to selectively excite propagating modes in the waveguide in a polarization-dependent way. This concept was originally exploited in particles near surfaces [14][83] and near optical fibers [84], but the effect is universal and can be easily extended to silicon photonic waveguides [85] (Fig 6(c)). Indeed, it provides a unique opportunity to introduce polarization control into silicon photonics via asymmetrically-placed antennas. While conventional nanoantennas radiate and receive light of a fixed polarization, employing the concepts of spin-direction locking we can design optical nanoantennas that sort the different polarization components of incident light along opposite guided paths. The concept can also be generalised to work with light polarizations other than circular, by using symmetry considerations, for instance enabling linear polarization sorting in silicon integrated nanoantennas [71,86] (Fig 6(d)). Notably, the inverse concept can also be exploited, as was done in plasmonics [14], to achieve optical nanoantennas with multiple inputs that can synthetize desired radiated polarizations [71]. When combined with the possibility of active phase and amplitude modulation in integrated waveguides, these multiple input/output nanoantennas could potentially lead to ultra-fast polarization synthesis and analysis, with applications in integrated ellipsometers [87], communications, quantum optics [88], and novel magnetic storage applications [89].

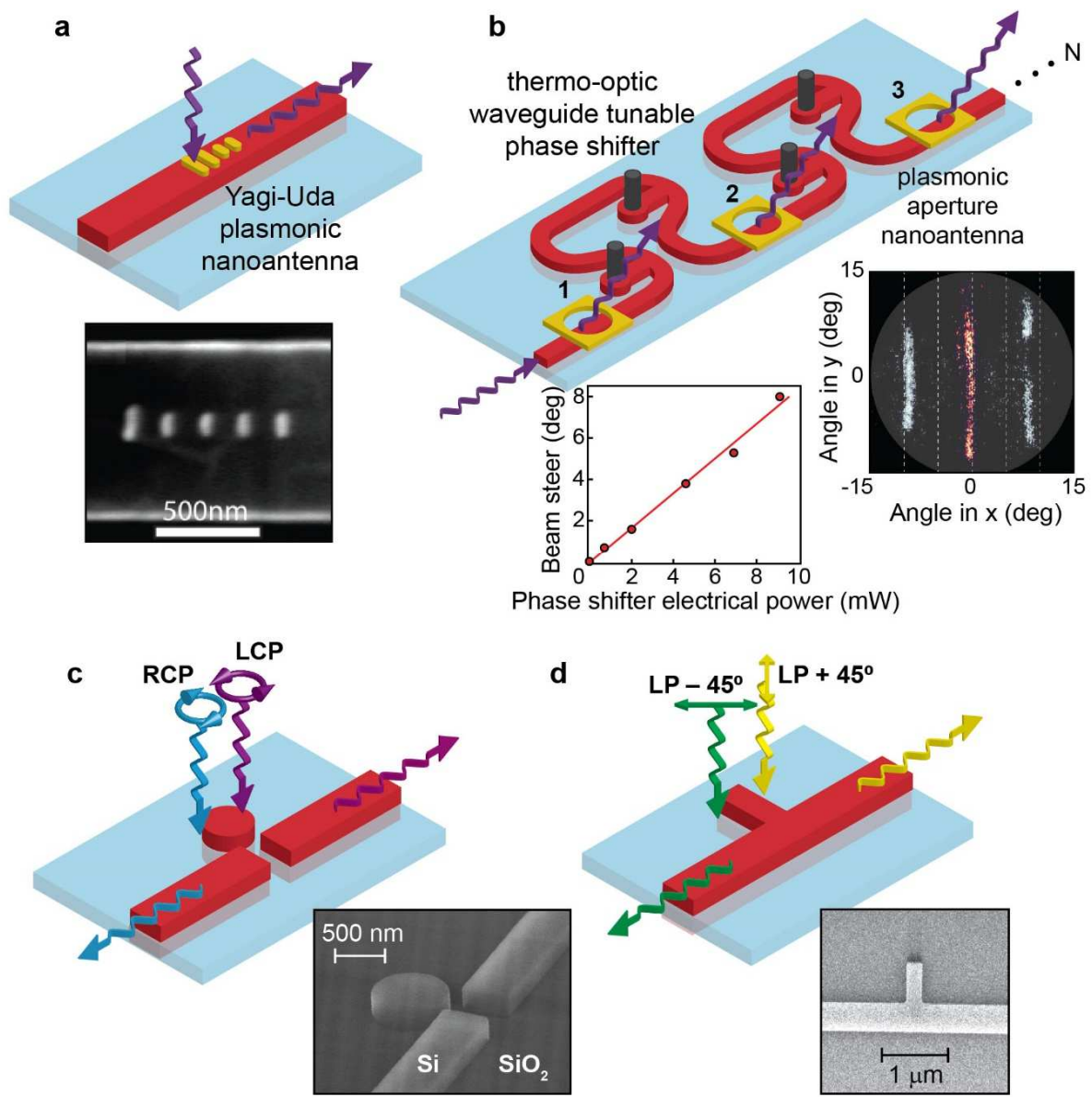

Figure 6. Experimental examples and measurements of nanoantennas on silicon PICs. (a) Plasmonic Yagi-Uda directional nanoantenna on a dielectric waveguide (inset: SEM image, Ref. [46]) (b) A plasmonic nanoantenna array on a dielectric waveguide is used for electrically-controlled steering of radiated light (inset: angular radiation pattern of the nanoantenna array and results on the electrical beam steering Ref. [76]) (c) Spin-orbit interaction in silicon photonics acting as a dual-input/output nanoantenna that can split the incident circular polarization components into opposite guided paths (inset: SEM image Ref. [90]). (d) Generalization of the idea of polarization splitting nanoantenna to other polarizations such as linear polarization (inset: SEM image Ref. [71])

## 4. Integration of metamaterials on silicon photonics

Ideally, metamaterials are 3D structures formed by subwavelength meta-atoms (typically plasmonic nanoresonators) whose response can be tailored to achieve any value of the effective electric permittivity and magnetic permeability. Being silicon photonics essentially a planar technology, it does not seem the most adequate choice for the implementation of 3D structures.

However, the situation changes if we consider metasurfaces, the 2D counterpart of metamaterials. In a metasurface, a 2D array of metallic nanoscatterers (they can be called nanoantennas), whose shape and size can be periodically modulated, is arranged upon a planar substrate [16]. Each nanoscatterer is tailored to display a certain amplitude and phase response for the scattered fields, enabling the control of a certain phase-front with a flat surface. Therefore, metasurfaces become a very powerful tool since they allow for a wide variety of functionalities traditionally realized using bulky optical elements (such as lenses, waveplates or splitters) on a surface occupying less than one wavelength. Interestingly, the use of metals is not mandatory: silicon, having a high index of refraction, can provide the required abrupt phase changes and reduced absorption losses, leading to the so-called all-dielectric metasurfaces [91]. Several flat optical devices using silicon metasurfaces have been demonstrated in the visible [92] (see Fig. 7(a)) and NIR [93] regimes, being their fabrication completely compatible with silicon technology.

Metasurfaces, as well as metamaterials, essentially perform in an out-of-plane configuration in which light is not guided in the silicon chip. But, remarkably, they can also be used in an in-plane geometry so that they interact with (and, therefore, process) guided light as in the case of the structures addressed in Sections 2 and 3. Indeed, some of the most relevant metamaterial devices and functionalities have been demonstrated within this geometry. For instance, let us consider the invisibility cloak [94], which is one of the most interesting functionalities that can be realized with metamaterials. A reduced version of such cloaks is the carpet cloak [95], which enables concealing a perturbation on a flat reflecting surface rather than a singular point. Interestingly, it relies on local variations of the electric permitivitty and therefore does not require magnetic activity. Carpet cloaks could be hence implemented by introducing local variations of the refraction index of a dielectric material. This approach was used to demonstrate a carpet cloak of on-chip guided light at telecom wavelengths on a SOI chip [96,97]. Local variations of the refractive index were achieved by changing the radius of the holes [97] (see Fig. 7(b)) or the density of pillars [96] in the cloaking region. Since no metals are involved, absorption losses are almost negligible. Moreover, the cloaking performance is broadband since it does not rely on resonant phenomena. Such demonstrations, together with the directional silicon nanoantennas described in the previous section, constitute good examples of phenomena that were first predicted or formulated for plasmonic waves in metals but that in silicon technology find a much better platform for experimental realization because of the extremely reduced losses.

More recently, hyperbolic metamaterials made of metal-dielectric pillars exhibiting an ENZ behaviour for light guided along silicon waveguides have been demonstrated [98] (Fig. 7(c)). Besides the exotic linear properties resulting from the lack of phase evolution, nonlinear properties are also strongly enhanced in ENZ materials as a result of the large field enhancement [99,100]. Therefore, ENZ metamaterials seem highly suitable candidates to implement all-optical ultrafast processing devices at low operating powers, application for which the integration in a silicon platform would result particularly relevant (large volume production, easy interconnection with optical fibers, and integration with electronics). Interestingly, a recent study analyses the possibility to attain all-optical switching with a 35% extinction rate using a hyperbolic metamaterial on top of a silicon waveguide [40]. Perhaps the performance in terms of extinction ratio and required switching power could be improved by using the embedded approaches previously addressed on Section 2.

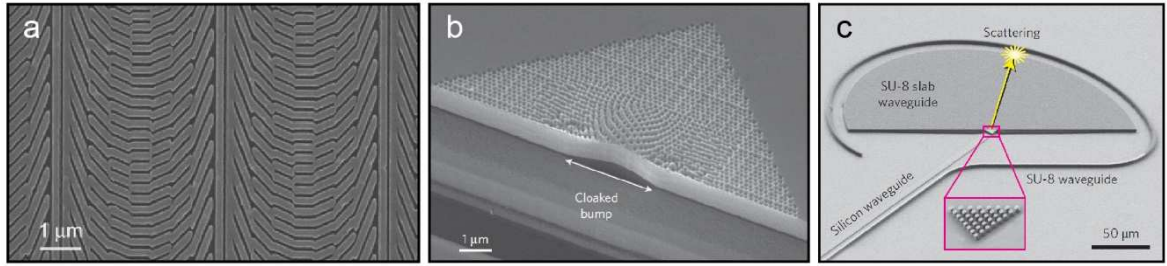

Figure 7. Implementation of metamaterials on silicon photonics: SEM images of (a) a metasurface operating as a blazed grating for out-of-plane visible light [92]; (b) a metamaterial carpet cloak created by digging holes on a SOI chip [97]; (c) an ENZ hyperbolic metamaterial made of metal-dielectric pillars (see inset) placed at the end of a silicon waveguide for measuring the refraction angle of guided waves [98].

## 5. CMOS compatible plasmonic materials for silicon photonics

A key issue to be considered in all new photonic structures to be integrated on a silicon PIC is their compatibility with CMOS processes used by silicon microelectronics industry. Such compatibility would ultimately allow for mass-manufacturing of advanced PICs at low production cost. Importantly, typical noble metals used in plasmonics and metamaterials, such as silver or gold, are not CMOS compatible, which is a major roadblock preventing the exploitation of the potential of plasmonics on silicon PICs [17]. Still, other metals such as copper (Cu) or aluminium (Al), which are fully CMOS compatible, can exhibit quite remarkable plasmonic properties in terms of moderate losses [101], LSPR linewidth [102], field confinement and, notably, nonlinear performance [33,103]. Figure 8 compares the transmission, reflection and field enhancement spectra of isolated metallic dipoles (formed by two nanostrips separated by a narrow gap) inserted in a silicon waveguide gap for various metals obtained via 3D FDTD simulations. Interestingly, the obtained response for Cu is similar to those obtained using Ag and Au. The response obtained with Al also shows a strong field enhancement in the plasmonic gap (hot-spot) though in this case the whole response is blue-shifted. It has to be mentioned that deleterious effects related to Al or Cu thin-layer deposition (notably surface oxidation and chemical stability) have not been considered in the simulations, which could result in an impairment of the plasmonic properties, unless techniques for surface protection are applied [104]. Therefore, a trade-off arises here between plasmonic performance and manufacturability, which should be properly considered in the development of practical applications. Besides Cu and Al, there are other materials that are widely employed in the silicon industry that can exhibit a remarkable plasmonic behaviour in the visible and near infrared regime. Indeed, recent papers have highlighted the exciting possibility of replacing noble metals by materials such as indium-tin-oxide (ITO) or titanium nitride (TiN) showing similar plasmonic properties to Au or Ag but with the added value of CMOS compatibility [105,106].

Another issue to be considered is feature resolution. Many plasmonic structures have been fabricated using processes such as focused-ion beam (FIB) or electron beam lithography (EBL) that enable high resolution in metallic features (< 10 nm with FIB) but low throughput. Getting resolution below 10 nm is highly relevant in boosting optical effects related to field localization, since in plasmonic gaps the field enhancement grows inversely to the gap size down to sizes of the order of 1 nm, where quantum and non-local effects come into play [107,108]. However, CMOS technology employs optical lithography to pattern large areas at high speed, which results in diffraction-limited best resolutions of the order of 20 nm [109,110]. Therefore, it is important to consider resolution limitations when thinking about practical silicon PICs containing plasmonic, nanoantennas or metamaterial elements.

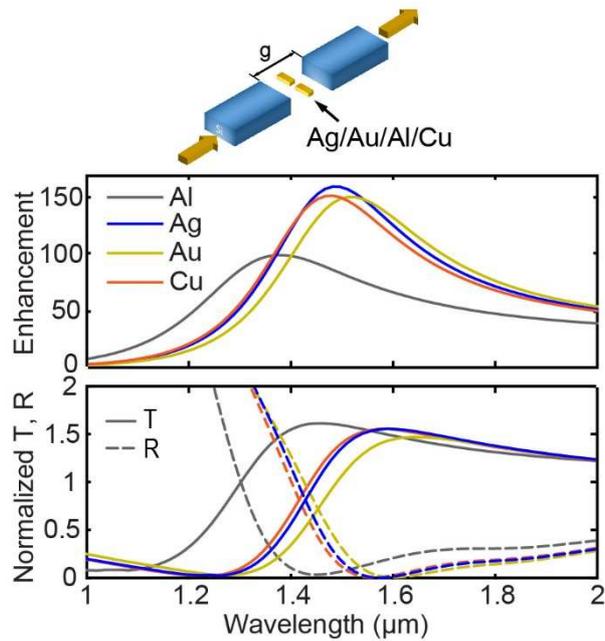

*Figure 8. Comparison of the response of plasmonic nanostructures made of different metals. Simulation results (3D FDTD using FullWave by Synopsys) of the intensity enhancement (top panel), transmission (bottom, solid line), and reflection (bottom panel, dotted line) provided by a metallic dipole (consisting of two 80x80x200 nm nanostrips spaced by a 60 nm gap inserted in a 400x250 nm silicon waveguide gap with 400 nm width) following the configuration depicted in Fig. 5. Several metals are considered: Al, Ag, Au and Cu. The waveguide is excited with the fundamental TE-like mode. The responses are obtained by monitoring the square of the transverse electric field at the center of the output waveguide (transmission), input waveguide (reflection) and metallic gap (intensity enhancement). The curves are normalized with those obtained in a configuration without the metallic nanostructure.*

## Summary

In this review we have discussed the advantages of bringing concepts unveiled in the fields of plasmonics, nanoantennas and metamaterials into silicon photonics, a suitable technological platform for mass-scale manufacturing of PICs. Plasmonic nanostructures and nanoantennas can be benefitted by the possibility of connecting multiple elements via lossless integrated waveguides, whilst silicon photonics could find a way towards further miniaturization of processing elements. In some cases, such as in nanoantennas, silicon can be used instead of the metal in order to reduce undesired absorption losses, though larger elements are obtained. If metals are to be employed, CMOS compatibility has to be addressed. Research in CMOS-compatible plasmonic materials such as ITO and TiN could be disruptive in the fields of plasmonics, nanoantennas and metamaterials since, first, the losses that now are a key roadblock in plasmonics could be reduced to a reasonable level; and second, enabling the fabrication of plasmonic components in CMOS foundries would allow research groups without (expensive) fabrication facilities to get samples for optical testing fabricated in a fabless fashion in silicon photonics foundries. In our opinion, bringing plasmonics, metamaterials and nanoantenna concepts into silicon photonics can be disruptive in many applications, including sensing, spectroscopy, polarization managing and high-speed data processing.

## Acknowledgements
A. M. and A. E.-S. acknowledge funding from contracts TEC2014-51902-C2-1-R and TEC2014-61906-EXP (MINECO/FEDER, UE) and, F. R.-F. acknowledges funding from EPSRC (UK).